# Improving institutional memory on challenges and methods for estimation of pig herd antimicrobial exposure based on data from the Danish Veterinary Medicines Statistics Program (VetStat)

**Nana Dupont[1], Mette Fertner[1], Anna Camilla Birkegård[2], Vibe Dalhoff Andersen[3], Gitte Blach Nielsen[1], Amanda Brinch Kruse[1], Leonardo Victor de Knegt[1,3]***

[1]Department of Veterinary and Animal Sciences, Faculty of Health and Medical Sciences, University of Copenhagen, Frederiksberg, Denmark.

[2]Section for Diagnostics and Scientific Advice, National Veterinary Institute, Technical University of Denmark, Frederiksberg, Denmark.

[3]Research group for Genomic Epidemiology, National Food Institute, Technical University of Denmark, Kongens Lyngby, Denmark.

**\* Correspondence:**
Leonardo Víctor de Knegt
leonardo@sund.ku.dk

## Abstract

With the increasing occurrence of antimicrobial resistance, more attention has been directed towards surveillance of both human and veterinary antimicrobial use. Since the early 2000s, several research papers on Danish pig antimicrobial usage have been published, based on data from the Danish Veterinary Medicines Statistics Program (VetStat). VetStat was established in 2000, as a national database containing detailed information on purchases of veterinary medicine. This paper presents a critical set of challenges originating from static system features, which researchers must address when estimating antimicrobial exposure in Danish pig herds. Most challenges presented are followed by at least one robust solution. A set of challenges requiring awareness from the researcher, but for which no immediate solution was available, were also presented. The selection of challenges and solutions was based on a consensus by a cross-institutional group of researchers working in projects using VetStat data. No quantitative data quality evaluations were performed, as the frequency of errors and inconsistencies in a dataset will vary, depending on the period covered in the data. Instead, this paper focuses on clarifying how VetStat data may be translated to an estimation of the antimicrobial exposure at herd level, by suggesting uniform methods of extracting and editing data, in order to obtain reliable and comparable estimates on pig antimicrobial consumption for research purposes.

# 1    Background

During the last two decades, antimicrobial resistance and responsible use of antimicrobials have become subjects of growing interest in the Public Health community. In the European Union alone, it is estimated that infections caused by antimicrobial-resistant bacteria are responsible for at least 25,000 human deaths per year, with societal cost of 1.5 billion Euros *per annum* (1-3). These figures are largely perceived as a consequence of imprudent antimicrobial use in both humans and animals (4,5). One of the key elements in understanding the development of antimicrobial resistance is a detailed knowledge of how antimicrobials are actually used. This has led international institutions, such as the World Health Organization and the European Council, to recommend increased surveillance of antimicrobial usage (3,6).

In Denmark, detailed data on all purchases of veterinary medicines have been collected since 2000 in the Danish Veterinary Medicines Statistics Program (VetStat). VetStat was originally designed with four aims: *"(1) to monitor veterinary usage of drugs in animal production; (2) to help practitioners in their work as herd advisors; (3) to provide transparency as a basis for ensuring compliance with rules and legislation and (4) to provide data for pharmaco-epidemiological research"* (7). While aims *(1)* to *(3)* have aided risk managers in decision-making processes and ensured compliance with legislative initiatives (8-10), aim number *(4)* has been the one to attract the most interest from the scientific community. For researchers, VetStat represents a valuable data source, containing detailed information on all purchases of veterinary medicine in Denmark. It is, therefore, tempting to use VetStat data to estimate antimicrobial exposure, as it: (i) requires no field work and is, consequently, cheaper than collecting primary data; (ii) is the closest secondary information to data on actual antimicrobial usage; (iii) has national coverage, including all Danish herds with production animals receiving prescription-only drugs; and (iv) stores data over time, allowing for retrospective, longitudinal studies to be performed. Consequently, VetStat data have been used in several studies and reports describing antimicrobial consumption in Danish pigs, quantified as kilograms (kg) of active ingredient (11), number of prescriptions (10) or number of Animal Defined Daily Doses (ADD) (12-15). However, VetStat stores data on the purchase of drugs, not treatments 7), so a series of assumptions and specific data management procedures are necessary to transform purchase data into an estimation of actual usage (Figure 1), and variation in data handling may lead to variation in results.

Challenges met while estimating actual antimicrobial use from VetStat records do not necessarily come from data errors and inconsistencies. In several occasions, such challenges originate from the way data is collected and how the system is structured. VetStat is a complex system, with a set of complementary tables, legal requirements and the possibility of data correction by users (pharmacies) and administrators (DVFA). For example, different drug batches must be recorded separately, even though they have been sold to fill the same prescription, resulting in two different records. Pharmacists have the possibility to retract sales retrospectively, by adding a subtractive record under the same prescription number, resulting in negative amounts when calculating consumption for that moment. Other challenges are, in fact, connected to data inconsistencies, but in complementary tables which are part of the system, baring no relation to the quality of the actual purchase entries They have, therefore, the potential to affect all sales extractions. Such system artifacts, along with others described in this paper, are static fixtures, occurring independently of herd selection, record inclusion/exclusion criteria, data cleaning approach and so forth. For these reasons, the focus of this paper was not on performing a quantitative data quality evaluation, nor detecting inconsistencies in one specific data extraction. The frequency of errors and inconsistencies in a dataset will vary, depending on the period covered, or on the time of data extraction. Additionally, the observed percentage of a given error may not necessarily reflect its true impact on exposure estimation. The existence of such fixtures may be less evident to researchers who have not worked intimately with VetStat data before, thus introducing a risk for misinterpretation of results. The structure of the VetStat database has been extensively described in previous papers, and the same can be said about the routines used to calculate ADDs and other standardized measures of antimicrobial usage (7,8). Although the establishment of uniform methods for extracting and editing data is necessary to obtain reliable and comparable estimates of antimicrobial consumption in pig herds, no publication has yet addressed the challenges encountered by researchers working on raw VetStat data. A documentation of such challenges is necessary to facilitate the use of VetStat data for future scientific studies and improve institutional memory on data management routines.

Hence, the objective of this paper is to present a set of challenges originating from static system fixtures in VetStat, which researchers must address when estimating antimicrobial exposure in Danish pig herds and, when possible, to offer at least one robust approach to deal with them. For individual researchers, the relevance of each challenge will depend on their own study objectives and methods. Challenges and solutions addressed in this paper may also prove useful for other countries considering the establishment of similar databases on veterinary drug use.

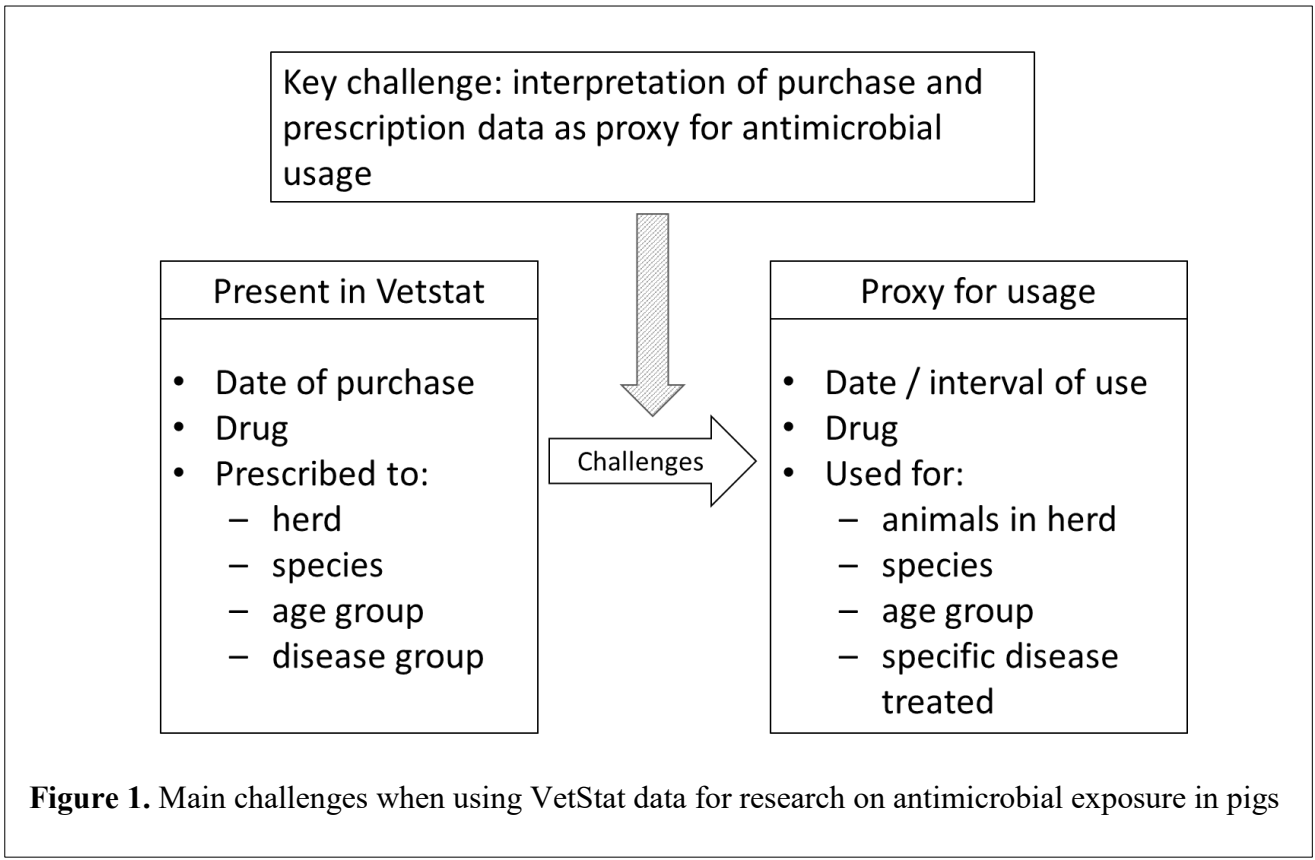

**Figure 1.** Main challenges when using VetStat data for research on antimicrobial exposure in pigs

## 2    Methods

### 2.1    Panel composition and selection of challenges and solutions

The present paper was developed by an inter-institutional panel of researchers. All members of the panel had two to five years of experience with VetStat data through involvement with previous research or advisory service.  Two members also had previous experience as veterinarians in Danish pig production farms. In an introductory meeting, all members put forward all challenges they had faced when trying to obtain robust exposure information using VetStat purchase data, and described how they chose to deal with those challenges. Challenges and solutions were discussed *in plenum* throughout a series of ten meetings, until a consensus on 12 challenges, with data-based solutions for eight, was reached. When necessary, further input was sought from DVFA employees maintaining or utilizing VetStat (cited here as personal communications), to assure consistency with the system's official description and working procedures. Although some challenges are related to data quality, no quantitative evaluations were performed, as the focus was placed on enabling researchers to critically assess their own individual datasets in relation to system artefacts, aside from the routine data cleaning and validation steps, which should, nonetheless, take place in every project.

### 2.2    Vetstat

VetStat is a national database containing detailed data on all purchases of prescription-only drugs for use in production animals. Consequently, VetStat data cover 100% of animal herds receiving medicine on prescription. In Denmark, antimicrobials for veterinary use can be purchased from pharmacies, veterinary practitioners or feed mills (7), and only with a valid prescription from a veterinarian (16). Therefore, all purchases of veterinary antimicrobials are recorded. VetStat is a relational database (Figure 2) on an Oracle platform, presently owned and managed by the Danish Veterinary and Food Administration (DVFA), a sub-department under the Ministry of Environment and Food of Denmark.

Each VetStat entry contains information on date of purchase, product identification code, amount of product purchased, identification code of the reporting pharmacy, veterinarian or feed mill, identification code of prescribing veterinarian, herd identification code,  as well as the age and disease groups targeted by the prescription (8). Records can be supplemented with information from additional tables, such as product trade names, concentrations, active ingredients, administration routes and dosages per kilogram of live animal weight (Figure 2). Pigs are divided into three age groups, with their corresponding standard weights, defined as the expected weight at treatment: (i) pre-weaning pigs, sows, boars and bred gilts (200 kg), (ii) weaners (15 kg) and (iii) finishers and non-bred gilts (50 kg). Six disease groups (indications) for pigs exist: 1- reproduction and urogenital system, 2- udder, 3- gastro-intestinal system, 4- respiratory system, 5- joints, limbs, hooves, skin and central nervous system and 6- metabolism, digestion and circulation (7).

The percentage of antimicrobial for use in pigs purchased from pharmacies, when compared to feed mills or veterinarians, has increased from 97.7% in 2002 to 99.9% in 2013 (17). It is, therefore, reasonable to assume that a solid exposure dataset can be obtained solely from pharmacy records. For that reason, and to limit the number of challenges described, this paper only includes challenges encountered when using data registered by pharmacies to estimate pig antimicrobial usage. A complete overview of the relevant VetStat tables used to estimate antimicrobial pig exposure based on pharmacy registrations is shown in Figure 2.

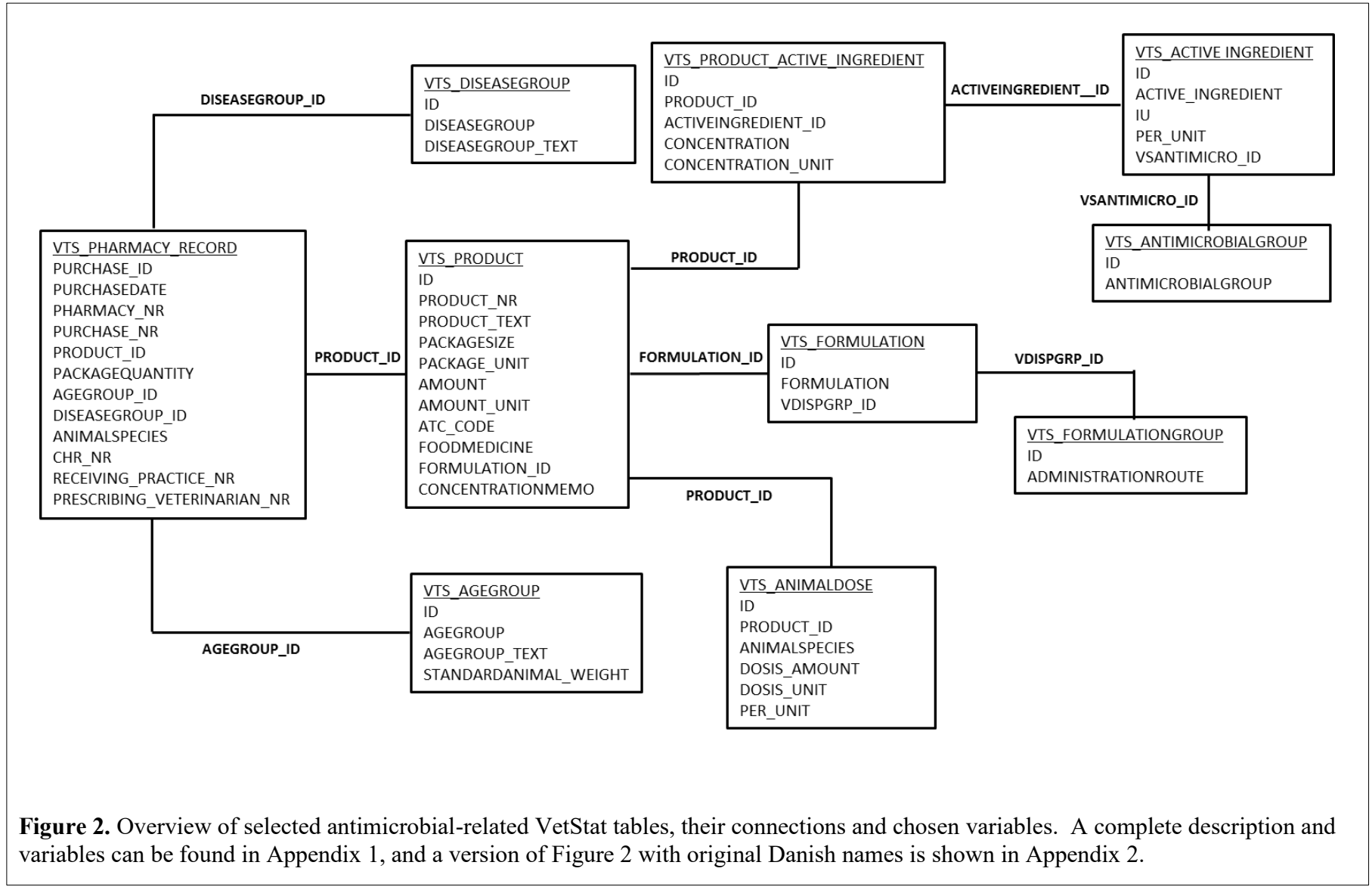

**Figure 2.** Overview of selected antimicrobial-related VetStat tables, their connections and chosen variables. A complete description and variables can be found in Appendix 1, and a version of Figure 2 with original Danish names is shown in Appendix 2.

## 2.3    Presenting challenges and solutions

Two types of challenges are presented in this paper. Section 3.1. contains eight challenges for which one or more robust data-based solutions could be proposed by the authors. Section 3.2. contains challenges about which researches should be aware, as they may have a significant impact on the interpretation of results and conclusions of a study, but no immediate solution was found. For consistency and clarity, the term "dataset" is used throughout the text in reference to data extracted from VetStat table "VTS_Pharmacy_Record", containing purchase records submitted by pharmacies. The term "table" is used when referring to instrumental VetStat tables containing complementary information, and which can be merged with purchase data (Figure 2).

## 3    Results

## 3.1    Challenges with proposed solutions

### 3.1.1  Inconsistencies between datasets extracted at different time points

Pharmacies upload data to VetStat on an ongoing basis. This means that historical information is generally stable, but data from current dates change continuously. For that reason, two extractions covering the same recent period may contain small data differences, if one of them is downloaded during a VetStat data update. Inconsistencies will also occur if there is a transfer fail during the extraction, or when manual corrections to single entries are performed by DFVA employees (E. Jacobsen, personal communication). No logs on such manual corrections are presently publicly available, and extracted datasets may, therefore, be non-reproducible.

*Solution:* The data should be checked for integrity of records. One indication of integrity is the presence of records in all months. Extracted datasets can also be validated by comparing the total amount of purchased antimicrobials (in kg of active ingredient) in an extraction with the numbers published on the DVFA website (18). If the researcher has access to a second extracted dataset, it is also possible to check for completeness of data by comparing the number of records in randomly selected parts of the two datasets. If the data used in a specific study seem at odds with data from similar populations or findings published in previous papers or reports, the dataset may be corrupted or incomplete. It should be kept in mind, when trying to reproduce historical series and validate data, that more than one dosage table has been used by official sources for calculation of ADDs throughout the years. One ADD has been defined as "*the assumed average maintenance dose per day for the main indication in a specified species*" (8,19).  The VetStat ADD table has been updated in 2014, so publications from DFVF before that year used a different table than the one available now. Similarly, DANMAP started using their own updated ADD, known as DADD, in 2012 (20), and reports published before this date used the pre-2014 VetStat ADD (21). Once a valid dataset is obtained, any further manual corrections should be written in a detailed log, together with the original date and time of extraction, to ensure transparency and reproducibility.

### 3.1.2 Negative entries

Entries with a negative amount of packages purchased occur when unused drugs are returned by the farmer, or corrections are made to original registrations. This poses a challenge for researchers summarizing data with time as a grouping variable, since negative entries can be registered in VetStat at any time following the original sale. According to DVFA working procedures, a negative entry may only be made by the pharmacy, if the corresponding positive entry is within the same calendar month. If corrections need to be made at a later time, they must be carried out by DVFA employees. If the original sale and the corresponding correction are located in different time blocks (as defined in each study design), the time difference between the two will result in an overestimation of consumption at the time of purchase, with a corresponding underestimation at the time of correction. The risk of isolating original records from their corresponding corrections increases, as time intervals summarizing consumption become smaller, e.g. when comparing weeks or months instead of years, and studies investigating smaller time intervals might suffer from biased estimates of antimicrobial consumption data. This challenge will also affect antimicrobial exposure assessments based on number of prescriptions, as each original entry with its retraction can be wrongly summarized as two prescriptions, if proper care is not taken.

*Solution*: It is recommended to investigate a buffer period around the actual study period, as it increases the likelihood of correctly pairing negative entries with their positive counterparts. In Denmark, the use of antimicrobials for oral medication is permissible up to a maximum of 35 days after the veterinarian has issued the prescription. For all other antimicrobials, the period is 63 days (purely finisher herds) or 50 days (all other herds) (16). All herds using antimicrobials for oral medication must, therefore, have monthly visits by a veterinarian. If only injectable antimicrobials are needed on a regular basis, visits may happen roughly every two months. Furthermore, as pharmacies transfer purchase data to VetStat at the end of each month, it is recommended to augment the buffer period by at least one more month, resulting in three months before and three months after the intended study period. Negative observations and their corresponding positive purchased amounts can be matched based on prescribing veterinarian, herd, pharmacy, product, disease group and date (a retraction must necessarily occur after or on the same date as the original purchase), and should be excluded from the final dataset. In some instances, the purchased and retracted amounts do not necessarily match, as not all the amount purchased is returned every time. In such cases, it is sensible to assume that positive differences reflect what was ultimately used in the herd, and should be maintained. Negative differences are likely related to errors, and should be handled accordingly to the researcher's data cleaning policy.

### 3.1.3 Incorrect identification number of prescribing veterinarian

Each entry in VetStat contains a number which identifies the prescribing veterinarian. These numbers are unique identifiers assigned by DVFA to all veterinarians in practice in Denmark. The veterinarian identification numbers are mostly used when prescription patterns, or the impact of individual veterinarians on herd antimicrobial exposure, are being investigated. Invalid veterinarian identification numbers, as well as valid, but wrongly identified identification numbers, may be present in VetStat records.

*Solution:* To reduce the risk of affiliating a veterinarian with a herd due to a wrongly typed veterinarian identification number, it is recommended to assign a minimum amount, or percentage of entries, in which the veterinarian in question appears as the prescribing veterinarian for the given herd. Herd affiliation for all practicing pig veterinarians can also be obtained from Vetreg, the national veterinary registry maintained by DFVA (http://www.vetreg.dk).

### 3.1.4 Incorrect animal species, age group or disease group identification code

As mentioned, VetStat classifies pigs according to one of three age groups: "pre-weaning pigs, sows, boars and gilts", "weaners" and "finishers". It is stated by law that one of these three approved age groups must be specified by the veterinarian on medicine prescriptions for pigs (16), and listed in the corresponding VetStat entry. However, prior to the Yellow Card legislation in 2010 (22), incidences occurred in which the veterinarian would write a prescription for "young pigs" (23). In such cases, the pharmacist had to assume in which of the three age groups the product should be registered for in VetStat.

Mismatches between animal species, age group and/or disease group may also occur due to typing errors at the pharmacies, as the pharmacy interface is not pre-coded. This allows for non-valid combinations, such as choosing animal species "cattle", age group "sows" and disease group "furunculosis" (coded as a fish disease) to occur in the same entry. According to Dupont (2016) (17), invalid or incorrect age groups corresponded to 1.5% of records in 2002, decreasing to 0.01% by 2013.

*Solution:* Since the introduction of the Yellow Card legislation, veterinarians and pharmacists have become more aware of the importance of correct age group terminologies and meticulous typing (23), as mistakes could result in sanctions being applied to the farmer. It is, therefore, expected that problems regarding invalid animal species, age and/or disease group identification codes will be less prevalent after 2010, as demonstrated for age group by Dupont (2016) (17). Still, when working with historical data, researchers should be aware that unexpected deviations in exposure for specific farms or age groups could be related to this issue.

Entries with mismatching animal species and/or age group identification code can be cross-validated with data from the Central Husbandry Register (CHR). The CHR is the national Danish database for registration of holdings and animals, which holds data on the number of pigs in all Danish herds, according to age group (24). In CHR, pigs are divided into weaners (7-30 kg), finishers (30 kg-slaughter) and breeding animals (200 kg). For all practical purposes, these correspond to the age groups used in VetStat. If a sale in VetStat refers to an animal species and age group, population data in CHR should indicate that both the species and age group exist in that herd. A lag time of six months to one year between actual change in types of age groups in a herd, and the corresponding CHR update, should be taken into account, as the number of animals must be updated twice per year for larger pig herds (>300 sows, >6000 weaners or >3000 finishers), and once per year for smaller herds (25).

### 3.1.5 Product identification for ADD calculations

In VetStat, all pharmacological products are identified by a unique product identification (Product ID), as well as a product number (Product NR) (Appendix 1). As an example, a 100 mL bottle of Penovet (300.000 IE/ml) and a 250 mL bottle of Penovet (300.000 IE/mL) will be identified by two different Product IDs and Product NRs. The difference between the two systems is that the Product ID corresponds to the "ID" variable in the Product Information table (Figure 2, Appendix 1), as an identifier for each row in the table. The numbers are unique and constant throughout the whole database, serving as a connection variable between VetStat tables. Product NR, on the other hand, is the product identifier adopted by DVFA as part of their pharmaceutical records. Product NRs are unique for active products. However, according to DVFA working procedures, discontinued products may have their Product NRs re-used in new, active ones. Working retrospectively on data covering several years, this may result in the same Product NR linking to two different products. This challenge is particularly relevant when calculating ADDs based on the DVFA dosages accessible at the VetStat website (https://VetStat.dk). Previously, ADDs were only calculated based on dosages from the VetStat table Animaldose (Figure 2), using the Product ID as identifier. However, when calculating ADDs using the DVFA dosages available online, the Product NR is used as identifier. When using retrospective data, this may result in wrongly applied standard dosages, for products which had their Product NRs re-used.

*Solution:* Discontinued products remain registered in the product dataset. It is, therefore, possible to detect Product NRs occurring more than once, and identify their correlated products with different trade names and Product IDs. Those products can be identified in the standard dosage table available at the VetStat website, and repeated Product NRs should be manually corrected to new, unique ones. It is important to assure that any changes in Product NR are performed identically, in both the website table and the table from the VetStat database, so identical Product NRs match completely for trade name, package size, concentration and administration route.

### 3.1.6 Doubling the purchased amount of combination products when calculating consumption in ADD for active ingredients

All active ingredients in a product are individually listed, with their corresponding concentrations, in the VetStat Product Active Ingredient table (Appendix 1). This means that combination products have one record (one row) for each active ingredient. The purchased amount of a combination product consisting of two active ingredients may therefore double, when purchase data from the pharmacies is merged with the table of active ingredients, since the original purchase entry will generate two nearly identical entries, with the full prescribed amount in each, where only active ingredients differ. When calculating consumption in ADDs, the amount of product purchased is used (Formula 1), and so the doubling of entries will lead to a doubling in the amount of sold combination products. This problem will also affect consumption estimates based on number of prescriptions, as all combination products will generate multiple rows in the final dataset. When quantifying antimicrobial purchase in kilograms of active ingredient, combination products do not give rise to concern, as the specific concentration of each ingredient is used to proportionally divide the purchased amount between them (Formula 2).

$$ADD \ = \frac{quantity\ of\ purchased\ product}{dosage\ per\ kg\ live\ weight * standard\ weight\ of\ corresponding\ age\ group}$$

**Formula 1**. Calculation of Animal Defined Daily Doses (ADD) based on standardized VetStat values on dosage per kg live weight and animal standard weights.

$$kilogram\ of\ active\ ingredient = \frac{quantity\ of\ purchased\ product * concentration\ (\frac{mg}{unit})}{1000000}$$

**Formula 2.** Calculation of kilogram of active ingredient based on values in VetStat

*Solution:* If there is a specific interest in calculating exposure to combination products, Anatomical Therapeutical Chemical (ATC) classification system codes (26) can be used to manually create an active ingredient variable. Combination products including a given active ingredient have different ATC codes than the isolated ingredient. This is particularly interesting when comparing the use of active ingredients as combinations and self-standing products. If active ingredients in a combination are to be assessed separately, it is possible to use the concentration variable in the Product Active Ingredient table (Appendix 1) to proportionally divide the purchased amount among them, so the sum of duplicated rows results in the original amount purchased. Comparisons between self-standing or combined use of ingredients can still be made when using this approach, as combinations will be found in rows where the calculated proportion will be less than 100%, making it possible to create a "combination indicator" variable. Both approaches, however, are passive of criticism, as they ignore the synergic effect of combination drugs, and assume that one whole ADD of each ingredient would be necessary for maintenance of treatment, even when they are combined. As no other option is currently available, these are acceptable solutions, but one should expect that specific ADDs for combination products are eventually made available by DFVF, taking into account the joint effect of two or more ingredients, when compared to their effect when used alone.

### 3.1.7 Date of purchase versus actual date of use of antimicrobials

VetStat records the date of purchase of a drug (Figure 1), but the actual date of use of the purchased drug is unknown. It may be assumed that an amount of a specific product is used in the interval between its initial purchase and the subsequent purchase of a product with identical composition and indication. However, as the interval between sales increases, so does the uncertainty regarding the actual time of usage. As mentioned in section 3.1.2. (negative entries), oral antimicrobials may be used up to 35 days after prescription, and other antimicrobials may be used up to 63 days after prescription in finisher herds, or 50 days in other herds. It is also possible for the veterinarian to re-validate a prescription, if the amount purchased before was not used. This re-validation is not registered in VetStat, and can be done any time during the after-prescription interval. Once re-validated, the 35, 63 or 50 days are reset, and the process can be repeated until the disease under treatment is no longer observed (16). Therefore, there might be significant differences between the date of purchase and the date of actual usage.

*Solution:* From a data point of view, long-term usage of purchased antimicrobials can be seen as sudden peaks in purchased amounts, followed by long periods with few or no prescriptions. It is important to remember that this behaviour can also be observed during outbreaks or seasonal diseases, but in the first case, it is likely that the number of peaks will be small throughout the study period, and in the second, peaks will normally be connected to season, occurring at more or less regular intervals. When investigating trends in antimicrobial consumption over time, particularly in periods smaller than one year, it is possible to calculate a moving average, by aggregating data in periods, and successively including a portion of the previously averaged amount in the calculation of the next period (e.g., when investigating trends between trimesters, include the last month of the previous trimester in the average calculation for the current one). This solution can smoothen unexpected differences between periods, but will not yield an estimate of the actual use in a specific month or period between two purchases. The second approach was described by Vigre et al. (2010) (27), who suggest calculating antimicrobial consumption between two dates of purchase, as the amount purchased on the first date divided by number of days until the next purchase. This solution allows for an estimation of consumption in individual weeks, months, or even days. When attempting to approximate a "per pig" estimation of treatment in different age groups in a period, it is recommended to expand the dataset or time interval, according to the probability of treatment, based on the expected age group turn-over. This recommendation is especially applicable when including finisher pig herds in the investigation, which tend to receive batches of pigs up to 13 weeks apart and are, therefore, expected to have a lower treatment frequency, when compared to weaner or sow herds, which will often have a higher turn-over of pigs.

### 3.1.8 Herd-level standardization of the amount of antimicrobials purchased

To allow between-herd comparisons, it is necessary to standardize the amount of purchased antimicrobials at herd level. For that, it is necessary to know the number of pigs potentially treated in a herd. Herd sizes are not available in VetStat datasets, and must be retrieved from other sources, such as the CHR. Changes in herd size may not be not instantly available, as farmers are only required to update their CHR information once or twice a year (see section 3.1.4.). It is also possible that the researcher only has access to a single or few CHR data extractions, not necessarily matched in time with their VetStat data. This may result in deviations between the herd sizes at the time when antimicrobials were purchased, and herd sizes according to the currently available data.

*Solution:* Population and purchase data will be more closely aligned, if monthly extractions of VetStat and CHR are paired. This enables the use of population data for more specific time blocks, increasing the likelyhood of detecting updates in CHR data. Alternatively, annual CHR extractions can be compared to each other, to identify and exclude herds with frequent large changes in the number of registered animals. If the herd sample size is relatively small, or the research project has sufficient resources, it is possible to manually obtain a consumption report for individual herds from the VetStat website, including the number of animals registered in the herd per month, according to the CHR.

When comparing antimicrobial exposure across production systems, it is recommended to collect supplementary information on CHR numbers which consist of more than one production unit, as different production units within the same CHR may use different production systems. If the study objectives allow it, it is recommended to exclude those CHR numbers, remembering to address any limitations this may cause to inferences and conclusions

## 3.2    Challenges requiring researcher awareness but without data-based solutions

For the following challenges, no data-based solutions could be identified. They cannot be solved just by using secondary data sources, and require supplementation from some type of primary data collected at herd level or interviews with farmers, veterinarians or pharmacists. An attempt can be made to obtain permission from DFVF to contact the specific pharmacy, prescribing veterinarian or herd for an in-depth trace-back of information. This can be very time-consuming, and relies on the feasibility of contacting the relevant parties.

### 3.2.1 Incorrect values in supplementary data

As mentioned earlier, aside from purchase data registered by the pharmacies, veterinarians and feed mills, VetStat also contains tables listing supplementary information, such as trade names, concentrations, package sizes, active ingredients, dosage per kg of live animal weight and administration routes (Figure 2, Appendix 1). These are manually entered into VetStat by DVFA employees. Typing errors in supplementary data are not very frequent, and are corrected as soon as they are detected (E. Jacobsen, personal communication), but have the potential to affect multiple registrations, from the day they occur. If the researcher finds, for example, a product with a dosage per kg of live animal weight which diverges from similar products (same active ingredient, concentration, indication and administration route), DVFA employees can be contacted to confirm or disprove the value in question.

### 3.2.2 Same prescription number used in more than one entry

This challenge affects investigations of prescribing behaviour in veterinarians, as well as antimicrobial exposure assessments based on number of prescriptions. When packages sold by the pharmacy originate from two different batches of drugs, both batch numbers must be recorded for future reference, resulting in two entries for the same prescription number. Information on product batch, however, is only available at the pharmacies, not VetStat. When packages of the same product are prescribed for treating different diseases within the same VetStat disease group (e.g. arthritis and meningitis: disease group for joints, limbs, hooves, skin and central nervous system), the veterinarian may write only one prescription, but one entry per disease must be generated, with the same prescription number. Finally, when packages of the same product are prescribed for different types of pigs within the same age group (e.g. treatment of tail bites in both unbred gilts and finishers), the pharmacy must, again, enter as many records as pig types being treated, for that same prescription. The described situations will result in the presence of multiple entries, in which only the sale identification number and, perhaps, the number of packages sold, differ.

### 3.2.3 Discrepancies between registered and actual treated disease groups

VetStat records should state the intended age group and disease group for the dispensed drug. However, according to the Danish Veterinary and Food Administration (2014) (23), a veterinarian may write a prescription for only the most prevalent disease groups under treatment, and recommend the use of the same drug for other disease groups requiring treatment at the same time. For studies focusing on disease groups, the Summary of Product Characteristics (28) can be checked to identify mismatches between official indications and disease groups registered in VetStat entries. Mismatches might indicate (i) records in which "herd diagnostic prescriptions" were used; (ii) records containing entry errors or (iii) records where the product was prescribed for off-label use. However, it is not possible, without tracing back the original prescription, to determine which one of those cases occurred. Additionally, instances in which a product was prescribed for an approved indication and used also for another disease, are not detectable. Still, instances in which the product indication and the registered use do not match may indicate herds for which a limitation to the drawing of conclusions must be recognized.

### 3.2.4 Difference in weight between sows and piglets for ADD calculations

As described by Stege et al. (2003) (7), antimicrobial purchase for sows and piglets are recorded under the same age group. When calculating use in ADDs for this age group, the standard weight of 200 kg (section 2.2.) is used in the denominator (Formula 1). This calculation assumes that all antimicrobials purchased were used for sows, which may result in an underestimation of antimicrobial exposure in ADDs in a herd, as it is expected that piglets are also treated. In a study conducted by the Danish Pig Research Centre (2015) (29), regular sows had an average of 17.9 piglets/litter (born with less than one kg), of which 11.7 reach weaning weight (around 7 kg). Therefore, the calculation error will vary, depending on the number of piglets present in a herd at the time the antimicrobials were used, as well as with their current weight. The disease group for the prescription, together with some knowledge on common treatment practice, can provide some insight on whether sows or piglets are being treated, but for diseases commonly affecting both, it is not possible to recover this information without recurring to the original herd or practitioner.

## 4    Discussion

VetStat is widely used as a data source in research (30-32), and it  is expected that, due to the increased focus on antimicrobial usage in production animals, this use will continue in future research projects. To aid future researchers in working with VetStat data as a proxy for antimicrobial exposure in Danish pigs, challenges commonly encountered during this process were described in this paper, with suggested approaches to deal with some of them. Many of those challenges are caused by entry retractions, typing errors or supplementary data problems. This underlines the fact that, regardless of the data origin, there is always a need to clean, check and validate data (33). Other challenges, however, were related to how the system is structured, how the data enters it, and how variables connect, when merging different datasets. Those were considered system artefacts, as opposed to data quality issues, and demand a certain knowledge about the database, in order to be properly detected and dealt with.

The focus of the present study was not on performing a quantitative data quality evaluation, which is why no percentages of detected problems in specific data extractions were presented. The frequency of errors and inconsistencies will vary between datasets, and the observed percentage of a given error may not necessarily reflect its true impact on exposure estimation. As an example, if only one product out of the 1,189 (0.08%) in the VetStat product information table has an ID number connecting it to a wrong standard dose, it will have no effect in a study using kg of active ingredient as exposure unit, but will have a large effect on the results, if ADDs are used as exposure unit, and the product is frequently purchased.

The key challenge when working with VetStat data is to transform register data on antimicrobial purchases into an estimate of the true exposure. Different clinical situations require different treatment lengths and number of daily medications. Furthermore, dosages used are adjusted according to the actual live weights of the animals. It is not possible to establish a system which can store all the information about each single treatment administered to each single pig in Denmark in a useful manner. In order to be accessible and useful for data analyses, such information must be summarized and standardized. During this process, the context and particularities of each specific antimicrobial usage is lost, as the original information is converted into a set of selected variables. In order to be viable, the system had to make a compromise to only include purchase data in VetStat, and to divide animals and diseases into set categories. The process of recording antimicrobial usage in Vetstat and trying to recover actual exposure information from it is, figuratively, like taking a high-resolution photography, reducing its definition to fit it into a document, and then attempting to use it as a full-sized picture again: after this process, a useful suggestion of the original contents remains there, but it is not possible to recover the lost pixels. For that reason, when attempting to re-create actual exposure from VetStat data, a series of assumptions must be made. Assumptions about time and method of usage become larger, as the desired level of detail in a study increases. As an example, it is safer to assume that a drug was used within a year, than within two weeks following its purchase. In a similar fashion, it is safer to estimate the overall consumption in an entire herd, when compared to estimating consumption in a specific age group, or to estimate the national consumption, as opposed to herd-level consumption.

Entries in VetStat refer to an entire age group in a herd, not a single animal. Consequently, information on how many treatments a single animal has received is not readily available, without directly contacting the herds. The traditional approach to estimate exposure on an average pig in a herd is to obtain herd size information from the CHR, and assume that (i) all pigs in the stated age group were treated; (ii) all purchased antimicrobials were used; and (iii) CHR data are correct and updated, both in terms of herd identification and pig population. Ideally, such assumptions should be validated through interviews, collection of primary data or other auxiliary datasets. The information for the three age groups is, then, summed by herd, generating an estimate of the exposure in a pig which has spent its whole life in a given herd. However, according to a report from the Danish Agriculture and Food Council, 43.3% of the Danish pig herds registered in 2014 were finisher herds, containing only slaughter pigs (34). This means that pigs in those herds originate from other herds, and much of their antimicrobial exposure history is missing from the "traditional" calculations. Given the observed differences in consumption patterns among herds (10,15,21), it is clear that pigs are subject to different exposure patterns during different parts of their lives, not only to what is calculated for the herd from where they are sent to slaughter. Analyses should, therefore, be appropriately adjusted for the potential movement of pigs between herds, as well as to the average number of days the pigs spend in the farrowing, nursery and finishing units. A method to estimate the life-time exposure to antimicrobials in an individual pig has been suggested by Andersen et al. in 2016 (35), taking into account the movement of pigs between herds but, ultimately, the most appropriate way of calculating exposure will be defined by the individual study objectives. For antimicrobial resistance studies using slaughterhouse samples, for example, antimicrobial treatments in the finisher herd may be of largest importance, as many types of resistance reduce with time from treatment (13). No two studies are identical, so, to decide which challenges may have implications, and which of the proposed solutions are the most appropriate, researchers should always critically assess their data in relation to their objectives.

The reader should keep in mind that this paper only addresses challenges encountered when working with data registered by pharmacies, as pharmacies presently report 99.9% of the Danish antimicrobial sale for pigs (17). Studies focusing on veterinarian clinical practices or prescription patterns may benefit from including data from the VetStat veterinarian registrations, just as studies focusing on mineral or vitamin supplementation are likely to find value in the feed mill registrations. Although some challenges presented in this paper will also find application in datasets based on veterinarian or feed mill registrations, working with these is likely to present a whole new set of unique challenges, such as large batch purchases from feed mills separated by substantial time intervals.

# 5    Conclusion

We conclude that, although VetStat is presently the best available source for secondary data on veterinary antimicrobial consumption in Denmark, there are several steps which must be taken, when converting register data into antimicrobial consumption or exposure estimates. The lack of information about the actual time and usage of a purchased antimicrobial product, as well as issues relating to system structure and the calculation of dose-based exposure, may bias estimations, and must be appropriately dealt with, or at least taken into account, before drawing any conclusions.

# 6    List of abbreviations

ADD: Animal Defined Daily Dose

ADDs: Animal Defined Daily Doses

ATC: Anatomical Therapeutical Chemical

CHR: Central Husbandry Register

DANMAP: The Danish programme for surveillance of antimicrobial consumption and resistance in bacteria from animals, food and humans

DVFA: The Danish Veterinary and Food Administration

**Competing interests**

The authors declare that the research was conducted in the absence of any commercial or financial relationships that could be construed as a potential conflict of interest.

**Funding**

No specific funding was received for this study, asides from the research means of the university employment of the authors.

**Acknowledgments**

The authors would like to thank Erik Jacobsen, Vibeke Frøkjær Jensen, Liza Rosenbaum Nielsen and Laura Mie Jensen for their indispensable advice and contribution of knowledge.

**References**

1. McEwen SA, Fedorka-Cray PJ. (2002). Antimicrobial use and resistance in animals. Clin Infect Dis 34 S3:93-106.
2. European Centre for Disease Prevention and Control. (2009). ECDC/EMEA Joint Technical Report. The bacterial challenge: time to react. European Centre for Disease Prevention and Control. http://ecdc.europa.eu/en/publications/Publications/0909_TER_The_Bacterial_Challenge_Time_to_React.pdf. Accessed 19 February 2016.
3. European Medicines Agency. (2011). Trends in the sales of veterinary antimicrobial agents in nine European countries (2005-2009). European Medicines Agency. http://www.ema.europa.eu/docs/en_GB/document_library/Report/2011/09/WC500112309.pdf. Accessed 14 February 2016.
4. Laxminarayan R, Duse A, Wattal C, Zaidi AK, Wertheim, HF, Sumpradit N et al. (2013). Antibiotic resistance—the need for global solutions. Lancet Infect Dis 13:1057-1098.
5. Hammerum AM, Larsen J, Andersen VD, Lester CH, Skytte TSS, Hansen F et al. (2014). Characterization of extended-spectrum β-lactamase (ESBL)-producing Escherichia coli obtained from Danish pigs, pig farmers and their families from farms with high or no consumption of third-or fourth-generation cephalosporins. J Antimicrob Chemoth 69:2650-2657.
6. World Health Organization. (2015). Global action plan on Antimicrobial Resistance. World Health Organisation. http://apps.who.int/iris/bitstream/10665/193736/1/9789241509763_eng.pdf?ua=1. Accessed 19 February 2016.
7. Stege H, Bager F, Jacobsen E, Thougaard A. (2003). VETSTAT—the Danish system for surveillance of the veterinary use of drugs for production animals. Prev Vet Med 57:105-115.
8. Jensen VF, Jacobsen E, Bager F. (2004). Veterinary antimicrobial-usage statistics based on standardized measures of dosage. Prev Vet Med 64:201-215.
9. Aarestrup FM, Jensen VF, Emborg H-D, Jacobsen E, Wegener HC. (2010). Changes in the use of antimicrobials and the effects on productivity of swine farms in Denmark. Am J Vet Res 71:726-733.
10. Jensen VF, de Knegt LV, Andersen VD, Wingstrand A. (2014). Temporal relationship between decrease in antimicrobial prescription for Danish pigs and the "Yellow Card" legal intervention directed at reduction of antimicrobial use. Prev Vet Med 117:554-564.

11. DANMAP. (2015). DANMAP 2014 - Use of antimicrobial agents and occurence of antimicrobial resistance in bacteria from food animals, food and humans in Denmark. The Danish Integrated Antimicrobial Resistance Monitoring and Research Programme. http://www.danmap.org/Downloads/Reports.aspx. Accessed 15 November 2015.

12. Emborg H-D, Vigre H, Jensen VF, Vieira A, Baggesen DL, Aarestrup FM. (2007). Tetracycline consumption and occurrence of tetracycline resistance in Salmonella typhimurium phage types from Danish pigs. Microb Drug Resist 13:289-294.

13. Vieira AR, Houe H, Wegener HC, Lo Fo Wong DMA, Emborg H-D. (2009). Association between tetracycline consumption and tetracycline resistance in *Escherichia coli* from healthy Danish slaughter pigs. Foodborne Pathog Dis 6:99-109.

14. Hybschmann, GK, Ersbøll A, Vigre H, Baadsgaard N, Houe H. (2011). Herd-level risk factors for antimicrobial demanding gastrointestinal diseases in Danish herds with finisher pigs: a register-based study. Prev Vet Med 98:190-197.

15. Fertner M, Boklund A, Dupont N, Enøe C, Stege H, Toft N. (2015). Weaner production with low antimicrobial usage: a descriptive study. Acta Vet Scand 57:38

16. Anonymous. (2015). Executive order 1362 of 30/11/2015 on veterinarians' use, distribution and prescription of medicinal products for animals (in Danish: BEK 1362 af 30/11/2015. Bekendtgørelse om dyrlægers anvendelse, udlevering og ordinering af lægemidler til dyr). Ministry of Environment and Food of Denmark. https://www.retsinformation.dk/Forms/R0710.aspx?id=175789. Accessed 17 February 2016.

17. Dupont N. (2016). Applications of VetStat data on pig antimicrobial usage. Opportunities, challenges and restrictive legislation. [PhD Thesis]. Copenhagen, Denmark. University of Copenhagen.

18. Danish Veterinary and Food Administration. (2016). VetStat - aktuelle antibiotikaopgørelser. Danish Veterinary and Food Administration - Ministry of Environment and Food of Denmark. http://www.foedevarestyrelsen.dk/Leksikon/Sider/VetStat.aspx. Accessed 15 February 2016.

19. DANMAP. (2010). DANMAP 2009 - Use of antimicrobial agents and occurence of antimicrobial resistance in bacteria from food animals, food and humans in Denmark. The Danish Integrated Antimicrobial Resistance Monitoring and Research Programme. http://www.danmap.org/Downloads/Reports.aspx. Accessed 20 November 2015.

20. DANMAP. (2013). DANMAP 2012 - Use of antimicrobial agents and occurence of antimicrobial resistance in bacteria from food animals, food and humans in Denmark. The Danish Integrated Antimicrobial Resistance Monitoring and Research Programme. http://www.danmap.org/Downloads/Reports.aspx. Accessed 29 November 2015.

21. Dupont N, Fertner M, Kristensen CS, Toft N, Stege H. (2016). Reporting the national antimicrobial consumption in Danish pigs: influence of assigned daily dosage values and population measurement. Acta Vet Scand 58:27. DOI: 10.1186/s13028-016-0208-5

22. Anonymous. (2010). Executive order 1319 of 01/12/2010 on special arrangements to decrease the antimicrobial consumption in swine herds (in Danish: BEK 1319 af 01/12/2010. Bekendtgørelse om særlige foranstaltninger til nedbringelse af antibiotikaforbruget i svinebesætninger). Ministry of Food, Agriculture and Fisheries of Denmark. https://www.retsinformation.dk/Forms/R0710.aspx?id=139431. Accessed 23 January 2016.

23. Danish Veterinary and Food Administration. (2014). Vejledning til sagsbehandling vedr. overskridelse af grænseværdier for antibiotikaforbrug. Danish Veterinary and Food Administration - Ministry of Food, Agriculture and Fisheries of Denmark. http://www.foedevarestyrelsen.dk/SiteCollectionDocuments/25_PDF_word_filer%20til%20down load/05kontor/Gult%20kort/140512%20Gult%20kort%20Vejledning%20ver%202.3.pdf. Accessed 15 February 2016.

24. Anonymous. (2014). Executive order 1383 of 15/12/2014 on tagging, registration and transport of cattle, swine, sheep or goats (in Danish: BEK 1383 af 15/12/2014. Bekendtgørelse om mærkning, registrering og flytning af kvæg, svin, får eller geder). Ministry of Food, Agriculture and Fisheries of Denmark. https://www.retsinformation.dk/Forms/R0710.aspx?id=166914. Accessed 13 February 2016.

25. Anonymous. (2013). Executive order 1237 of 30/10/2013 on registrations of herds in CHR (in Danish: BEK 1237 af 30/10/2013. Bekendtgørelse om registering af besætninger i CHR). Ministry of Food, Agriculture and Fisheries of Denmark. https://www.retsinformation.dk/Forms/R0710.aspx?id=158819. Accessed 17 February 2016.

26. World Health Organization. (2011). ATC - Structure and principles. WHO Collaborating Centre for Drug Statistics Methodology. http://www.whocc.no/atc/structure_and_principles/. Accessed 11 January 2016.

27. Vigre H, Dohoo IR, Stryhn H, Jensen VF. (2010). Use of register data to assess the association between use of antimicrobials and outbreak of Postweaning Multisystemic Wasting Syndrome (PMWS) in Danish pig herds. Prev Vet Med 93:98-109.

28. Danish Medicines Agency. (2016). Summaries of product characteristics. https://laegemiddelstyrelsen.dk/en/sideeffects/find-medicines/summaries-of-product-characteristics. Accessed 15 November 2016.

29. The Danish Pig Research Centre. (2015). Annual Report 2014. Danish Agriculture and Food Council. http://www.pigresearchcentre.dk/~/media/Files/PDF%20-%20Aarsberetning%20VSP%20English/%C3%85rsberetning%202014_UK.pdf. Acessed 10 November 2016.

30. Pedersen K, Jensen H., Finster K, Jensen VF, Heuer OE. (2007). Occurrence of antimicrobial resistance in bacteria from diagnostic samples from dogs. J Antimicrob Chemoth 60:775-781.

31. Vigre H, Larsen P, Andreasen M, Christensen J, Jorsal S. (2008). The effect of discontinued use of antimicrobial growth promoters on the risk of therapeutic antibiotic treatment in Danish farrow-to-finish pig farms. Epidemiol Infect 136:92-107.

32. Fertner M, Sanchez J, Boklund A, Stryhn H, Dupont N, Toft N. (2015). Persistent Spatial Clusters of Prescribed Antimicrobials among Danish Pig Farms – A Register-Based Study. PloS ONE 10 (8), e0136834.

33. Emanuelson U, Egenvall A. (2014). The data – Sources and validation. Prev Vet Med 113:298-303.

34. Danish Agriculture and Food Council. (2014). Statistics 2014 - Pigmeat. Danish Agriculture and Food Council. http://lf.dk/Tal_og_Analyser/Aarstatistikker/Statistik_svin/Statistik_svin_2014.aspx. Accessed 10 December 2015.

35. Andersen VD, de Knegt LV, Munk P, Jensen MS, Agersø Y, Vigre H. (2016). The effect of antimicrobial usage on occurrence of resistance genes in fecal samples from slaughter pig batches. The Annual Meeting of the Society for Veterinary Epidemiology and Preventive Medicine 2016. Elsinore, Denmark.

**Appendix 1**. Names and descriptions of selected VetStat tables and their variables.

| Dataset | | Variables | | |
|---|---|---|---|---|
| **Name** | **English name and description** | **Name** | **Name in English** | **Description(a)** |
| VTS_APO_MED_REG | VTS_PHARMACY_RECORD Records of sales performed by pharmacies | ID | PURCHASE_ID | Unique identification number of purchase |
| | | UDLEVERINGSDATO | PURCHASEDATE | Date in which the purchase was made. |
| | | APOTEK_NR | PHARMACY_NR | Identification number of the pharmacy where the purchase was made. |
| | | EKSPEDITIONS_NR | PURCHASE_NR | Number of the transaction in the pharmacy system. |
| | | EKSPEDITIONS_TYPE | PURCHASETYPE | Originates from the pharmacy system. Marks records to be transferred to Vetstat. All values are filled as DI (large volumes sold to vets) or ID (other veterinary uses). |
| | | VARE_ID | PRODUCT_ID | Linking variable with dataset VTS_PRODUCT. Works as unique identifier number for products. |
| | | PAKNINGSANTAL | PACKAGEQUANTITY | Number of packages of the product being sold. Further detailing on the quantity is obtained by using variables PACKAGESIZE and AMOUNT, from dataset VTS_PRODUCT. |
| | | ALGR_ID | AGEGROUP_ID | Linking variable with VTS_AGEGROUP. |
| | | ORGR_ID | DISEASEGROUP_ID | Linking variable with dataset VTS_DISEASEGROUP. |
| | | DYREART | ANIMALSPECIES | Number code of the species of the animal for which the product was prescribed. |
| | | CHR_NR | CHR_NR | Identification number of the herd for which the product was prescribed, as found in the CHR |
| | | MODT_PRAKSISNR | RECEIVING_PRACTICE_NR | When a product is sold to a veterinarian for in situ use, this variable is filled in place of the herd number, otherwise it is left blank. |
| | | RECEPT_UDST_AUT_NR | PRESCRIBING_VETERINARIAN_NR | Unique identifier of a veterinarian assigned by DVFA. |
| | | KORREKTIONSCODE | CORRECTION_CODE | Not in use by Vetstat. Used by the pharmacy to assign corrections to entries. |
| | | FVREG_ID | FVREG_ID | Not in use. Originally referred to DVFA Regions. |

In: Dupont N, Fertner M, Birkegård AC, Andersen VD, Nielsen GB, Kruse AB, de Knegt LV. 2017. Improving institutional memory on challenges and methods for estimation of pig herd antimicrobial exposure based on data from the Danish Veterinary Medicines Statistics Program (VetStat).

| | | | | |
|---|---|---|---|---|
| VTS_ORDINATIONSGRUPPE | VTS_DISEASEGROUP<br>Diagnostic group of the disease for which the drug has been prescribed | ORDINATIONSTEKST | DISEASEGROUP_TEXT | Diagnostic group in text. Describes the functional system for which the drug is being prescribed. E.g. "respiratory tract disorders" or "urogenital and reproductive tract disorders". |
| | | ORDINATIONSGRUPPE | DISEASEGROUP | Diagnostic group coded as numbers. |
| | | DWORDINATIONSGRUPPE | DWDISEASEGROUP | Alphanumeric field with the diagnostic groups' textual description and number codes merged together. |
| VTS_ALDERSGRUPPE | VTS_AGEGROUP<br>Age groups or production categories of the animals for which the drug has been prescribed | ALDERSGRUPPE | AGEGROUP | Animal age groups coded as numbers. E.g. 55 for sows/piglets, 56 for weaners, 57 for slaughter pigs. |
| | | ALGR_TEKST | AGEGROUP_TEXT | Textual description of the age groups. E.g. "Sows/piglets", "weaners", "slaughter pigs". |
| | | DWALGR_TEKST | DWAGEGROUP_TEXT | Alphanumeric field with the age groups' textual description and number codes merged together. |
| | | STANDARDDYR_VAEGT | STANDARDANIMAL_WEIGHT | Standard weight for a given age group. |
| Instrumental variables | Present in most or all datasets | ID | ID | Unique identifier or a record (row) in a dataset. Serves as linking variable in other datasets. E.g. ID numbers from VTS_PRODUCT appear in other sets as PRODUCT_ID. |
| | | DATO_OPRET | DATE_OPENING | Date in which a record was entered in the dataset. |
| | | DATO_RET | DATE_CORRECTION | Last date a correction was made to a row. |
| | | BRUGER_OPRET | USER_OPENING | Initials of the user who entered the row in the dataset. |
| | | BRUGER_RET | USER_CORRECTION | Initials of the user who last made a correction to the row. |

| Dataset | | Variables | | |
|---|---|---|---|---|
| **Name** | **English name and description** | **Name** | **Name in English** | **Description(a)** |
| VTS_VARE | VTS_PRODUCT<br>Product information at label level | VARE_NR | PRODUCT_NR | Identification number of the product. Not unique. Values can be re-used for new products when old ones are discontinued. |
| | | VARE_TEKST | PRODUCT_TEXT | Name of the product described in that row. |
| | | PAKNINGSTOERRELSE | PACKAGESIZE | Number of sub-units in one package. E.g. one package may contain six flasks, or eight applicators, etc. |
| | | ENHED_PAKNING | PACKAGE_UNIT | Unit of the sub-units contained in the package. E.g. bottle, applicator, envelope. |
| | | MAENGDE | AMOUNT | Quantity of the preparation contained in each sub-unit. E.g. a bottle with 500 mL will have the value "500". |
| | | ENHED_MAENGDE | AMOUNT_UNIT | Unit of the preparation contained in the sub-unit. E.g. mL, grams, etc. |
| | | ATC_KODE | ATC_CODE | Anatomical Therapeutic Chemical (ATC) Classification System code for the active ingredients or combinations in the preparation (b). |
| | | DATO_TIL | DATE_TIL | Date until which prescriptions of a registered product will be accepted in Vetstat. Normally filled with the default value "31122099". |
| | | FODERMEDICIN | FOODMEDICINE | Dichotomous variable. If the product is supposed to be mixed in food, the value "1" is typed in, otherwise, "0". |
| | | VDIS_ID | FORMULATION_ID | Linking variable with dataset VTS_FORMULATION. |
| | | STYRKENOTAT | CONCENTRATIONMEMO | Total solid concentration of active ingredient found in the product, with units for the numerator and denominator. E.g. "150 mg/ml" or "100 IU/ml". |
| | | IE_OMREGN | IU_CONVERSION | Numerical value of the preparation quantity in International Units (IU). |
| | | ML_OMREGN | ML_CONVERSION | Numerical value of the preparation quantity in mL. |
| | | DS_OMREGN | DS_CONVERSION | Numerical value of the preparation quantity in doses. |
| | | REFP_ID | REFP_ID | Assigns when the pharmacy hands out another, but similar product to the one prescribed. Mostly used by the cattle database (Kvægdatabasen). |

| VTS_VARE_AKTIVT_STOF | VTS_PRODUCT_ACTIVE_INGREDIENT — Active ingredient and concentration information, as connected to each product | VARE_ID | PRODUCT_ID | Linking variable with dataset VTS_PRODUCT. |
|---|---|---|---|---|
| | | AKTS_ID | ACTIVEINGREDIENT_ID | Active ingredients present in the preparation coded as numbers. Links with variable ID in dataset VTS_ACTIVE_INGREDIENT for textual values. |
| | | STYRKE | CONCENTRATION | Numerical value of the total solid concentration of one active ingredient in the product. |
| | | ENHED_STYRKE | CONCENTRATION_UNIT | Unit of the solid part of the concentration (numerator). E.g. gram, mg, mL. |
| | | ENHED_PR_ENHED | UNIT_PER_UNIT | Unit of the concentration denominator, as related to the product administration form. E.g., mL (solutions), mg (powders), tablet, applicator, etc. |
| | | IE | IU | Numerical value of the solid part of the ingredient in IU. Variable not used. Instead, see variable IE in dataset VTS_ACTIVE_INGREDIENT. |
| | | IE_PR_ENHED | IU_PER_UNIT | Unit of the concentration denominator. Variable not used. Instead, see variable PER_UNIT in dataset VTS_ACTIVE_INGREDIENT. |
| VTS_AKTIVT_STOF | VTS_ACTIVE_INGREDIENT — Active ingredient information, as related to international units and antibiotic groups | AKTIVT_STOF | ACTIVE_INGREDIENT | Textual active ingredient name. |
| | | IE | IU | Numerical value for the numerator of the active ingredient concentration in IU. |
| | | PR_ENHED | PER_UNIT | Unit of the concentration denominator in IU, as related to the product administration form. E.g., mL (solutions), mg (powders), tablet, applicator, etc. |
| | | VSANTIBIO_ID | VSANTIMICRO_ID | Antimicrobial group, coded as numerical values. Linking variable with dataset VTS_ANTIMICROBIALGROUP. |

| VTS_ANTIBIOTIKAGRUPPE | VTS_ANTIMICROBIALGROUP Antimicrobial groups, textual values | ANTIBIOTIKAGRUPPE | ANITMICROBIALGROUP | Textual names of antimicrobial groups to which the active ingredients belong. |
|---|---|---|---|---|
| VTS_DISPENSERING | VTS_FORMULATION Administration routes and therapeutic forms | DISPENSERINGSFORM | FORMULATION | Product formulation. E.g. tablet, applicator, nose spray. |
| | | VDISPGRP_ID | VDISPGRP_ID | Linking variable with dataset VTS_FORMULATIONGROUP. Administration route groups (oral, parenteral, others), coded as numerical values. |
| VTS_DISPGRUPPE | VTS_FORMULATIONGROUP Administration routes and therapeutic forms, grouped | DISPGRUPPE | ADMINISTRATIONROUTE | Textual names of administration route groups. E.g. "oral", "parenteral". |

| | | VARE_ID | PRODUCT_ID | Linking variable with dataset VTS_PRODUCT. |
|---|---|---|---|---|
| VTS_DYREEKVIVALENT | VTS_ANIMALDOSE | DYREART | ANIMALSPECIES | Number code of the species of the animal for which the product was prescribed. |
| | | DOSIS_MAENGDE | DOSIS_AMOUNT | Numerical value for the amount of product preparation corresponding to one treatment dose per kg live weight.. |
| | | ENHED_DOSIS | DOSIS_UNIT | Unit for the amount of product (numerator) in the dose. |
| | | PR_ENHED | PER_UNIT | Unit of the animal live weight (denominator) in the dose, normally in kg. Pre-dosed products have the unit "ds." or "tablet" in the numerator and denominator. |
| Instrumental variables | Present in most or all datasets | ID | ID | Unique identifier of a record (row) in a dataset. Serves as linking variable in other datasets, named as an abbreviation of the dataset of origin, followed by "_ID". As an example, ID numbers from VTS_PRODUCT appear in other sets as PRODUCT_ID. |
| | | DATO_OPRET | DATE_OPENING | Date in which a record was entered in the dataset. |
| | | DATO_RET | DATE_CORRECTION | Last date a correction was made to a row. |
| | | BRUGER_OPRET | USER_OPENING | Initials of the user who entered the row in the dataset. |
| | | BRUGER_RET | USER_CORRECTION | Initials of the user who last made a correction to the row. |

(a) Descriptions and code interpretation based on the legislation for reporting of information for drug statistics, by the Danish Ministry of Health. Original title "Bekendtgørelse om indberetning af oplysninger til lægemiddelstatistik", Sundheds- og Ældreministeriet. Available from https://www.retsinformation.dk/pdfPrint.aspx?id=135928.

(b) World Health Organization Collaborating Centre for Drug Statistics Methodology. WHOCC ATC/DDD Index 2015. Available from http://www.whocc.no/atc_ddd_index/ .

**Appendix 2.** Overview of selected antimicrobial-related VetStat tables, their connections and chosen variables.

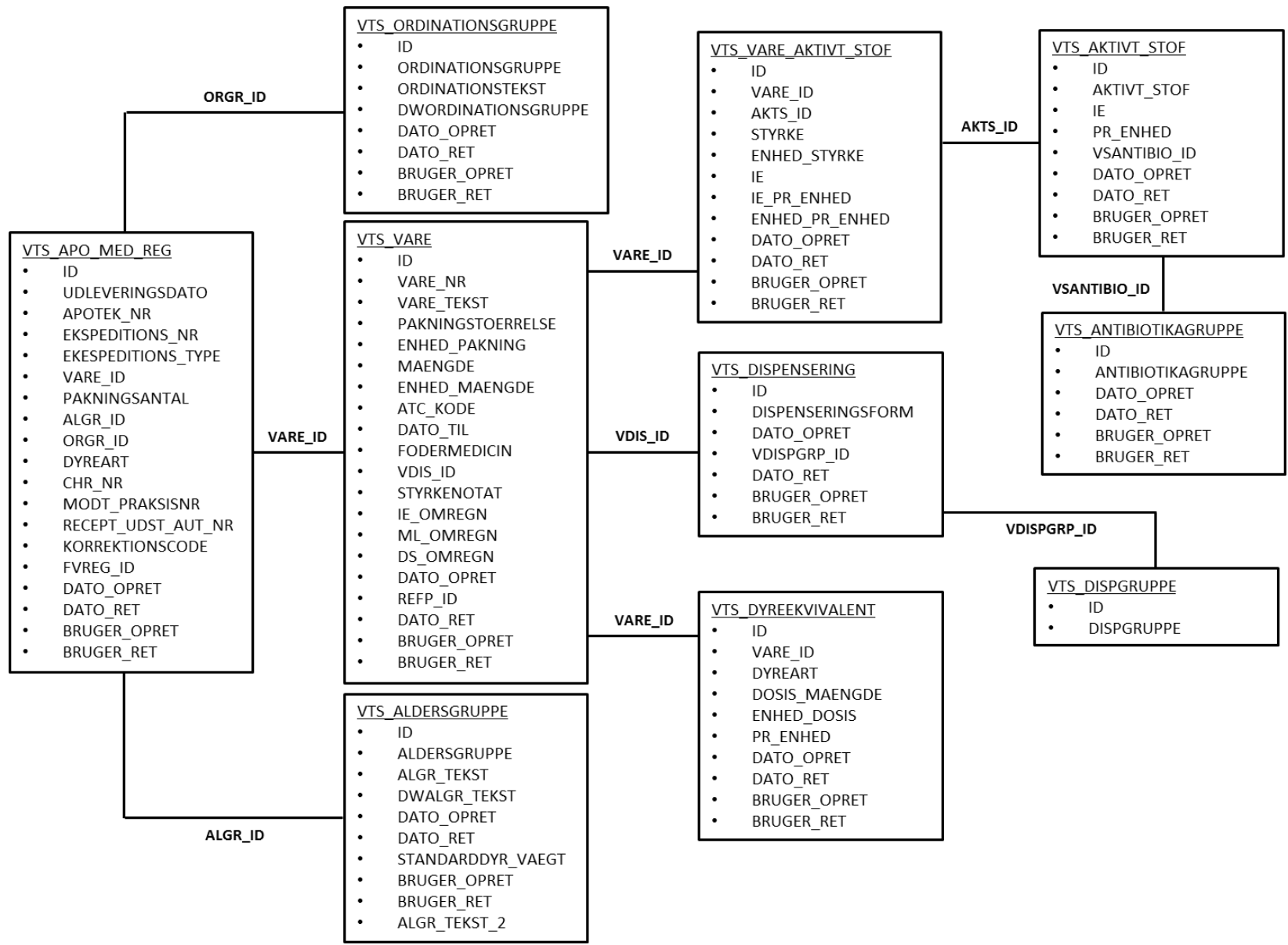

In: Dupont N, Fertner M, Birkegård AC, Andersen VD, Nielsen GB, Kruse AB, de Knegt LV. 2017. Improving institutional memory on challenges and methods for estimation of pig herd antimicrobial exposure based on data from the Danish Veterinary Medicines Statistics Program (VetStat).